\documentstyle[prl,aps,multicol,epsfig,rotating]{revtex}

\renewcommand{\narrowtext}{\begin{multicols}{2}}
\renewcommand{\widetext}{\end{multicols}}

\begin{document}
\draft

\newcommand{\lsim}   {\mathrel{\mathop{\kern 0pt \rlap
  {\raise.2ex\hbox{$<$}}}
  \lower.9ex\hbox{\kern-.190em $\sim$}}}
\newcommand{\gsim}   {\mathrel{\mathop{\kern 0pt \rlap
  {\raise.2ex\hbox{$>$}}}
  \lower.9ex\hbox{\kern-.190em $\sim$}}}
\def\be{\begin{equation}}
\def\ee{\end{equation}}
\def\ba{\begin{eqnarray}}
\def\ea{\end{eqnarray}}
\def\d{{\rm d}}

\newcommand{\promille}{%
    \relax\ifmmode\promillezeichen
          \else\leavevmode\(\mathsurround=0pt\promillezeichen\)\fi}
  \newcommand{\promillezeichen}{%
    \kern-.05em%
    \raise.5ex\hbox{\the\scriptfont0 0}%
    \kern-.15em/\kern-.15em%
    \lower.25ex\hbox{\the\scriptfont0 00}}

\def\ap{\approx}
\def\eff{{\rm eff}}
\def\P{{\mathcal P}}
\def\theta{\vartheta}
\def\epsilon{\varepsilon}

\def\tmin{t_{\rm min}}
\def\zmin{z_{\rm min}}
\def\zmax{z_{\rm max}}

\title{Electroweak jet cascading in the decay of superheavy particles}

\author{V.~Berezinsky$^{1}$, M.~Kachelrie{\ss}$^{2}$, and
        S.~Ostapchenko$^{3,4}$} 
\address{$^1$INFN, Laboratori Nazionali del Gran Sasso, I--67010 Assergi (AQ)\\
         $^2$Max-Planck-Institut f\"ur Physik (Werner-Heisenberg-Institut), 
             D--80805 M\"unchen\\
         $^3$Institut f\"ur Experimentelle Kernphysik,
             Universit\"at Karlsruhe, D--76021 Karlsruhe\\
         $^4$Moscow State University, Institute of Nuclear Physics,
             199899 Moscow, Russia } 

\date{May 17, 2002 --- MPI-PhT 2002-20}

\maketitle
\begin{abstract}
We study decays of superheavy particles $X$ into leptons. We show that
they initiate cascades similar to QCD parton jets, if
$m_X\gsim 10^6$~GeV. Electroweak (EW) cascading is studied and the
energy spectra of the produced leptons are calculated in the framework
of a broken SU(2) model of weak interactions. 
As application, important for the $Z$-burst model for ultrahigh energy
cosmic rays, we consider decays of superheavy particles coupled on 
tree-level only to neutrinos and derive stringent limit for these 
decays from the observed diffuse extragalactic $\gamma$-ray flux. 
\end{abstract}

\narrowtext
%
%
A characteristic feature of high energy deep-inelastic scattering, 
$e^+e^-$-annihilation and decays of superheavy particles is the
cascading of QCD partons. Despite the smallness of the QCD coupling
$\alpha_3$, cascading occurs because the probability of
parton splitting is enhanced by large logarithms for soft or collinear
parton emission. Recently, similar logarithms were found to dominate
also the EW radiative corrections at the TeV scale and above~\cite{EW}.
Evolution equations for these corrections, similar
to the DGLAP equations~\cite{DGLAP1,DGLAP2} in QCD but valid for a
spontaneously broken theory, were derived in~\cite{CCC}. 
In this Letter, we want to draw attention to decays of particles much
heavier than the EW scale: if these particles decay or annihilate, 
then the effects discussed above result in particle cascades   
which proceed through all U(1)$\times$SU(2)$\times$SU(3) interactions
with couplings $\alpha_1$, $\alpha_2$ and $\alpha_3$ enhanced
by logarithmic terms. Particular attention is given to 
EW Lepton-Boson (EWLB) cascades developing without QCD partons.

We shall illustrate the development of EW cascades by the example
of a superheavy $X$-particle with mass $m_X \lsim m_{\rm GUT}$ decaying
into leptons, which  has an interesting physical application to the
$Z$-burst model (see below). Let us discuss first the case when the
$X$-particle has on tree-level only couplings to lepton pairs 
$\bar ll$ and consider the decay mode $X\to\bar\nu\nu$. 
For $m_X\gg m_Z$, the mass of the $Z$ boson is  
negligible compared to the available momentum flow $Q^2\leq m_X^2/4$. 
Then, soft and collinear singularities generate large logarithms
$\ln^2(m_X^2/m_Z^2)$, which can compensate the 
smallness of $\alpha_2/(2\pi)$. 

To show the existence of the cascade, we consider 
the ratio $R=\Gamma(X\to\bar\nu\nu Z)/\Gamma(X\to\bar\nu\nu)$. 
Neglecting terms finite in the limit $m_Z\to 0$, it is given by
\be \label{R}
R = \frac{\alpha_2}{8\pi c_W^2}
         \left( \ln^2\epsilon + 3\ln\epsilon  +\ldots \right) ,
\ee
where 
$\epsilon=(m_Z/m_X)^2$ and $c_W^2=\cos^2\theta_W$.
The $\ln^2\epsilon$ term---which 
compensates the small coupling---is typical for 
soft and collinear
singularities in the emission of vector particles~\cite{wegr}. 
For $m_X\sim 10^6$~GeV, the decay probability into $\bar\nu\nu$
and $Z\bar\nu\nu$  
becomes comparable, $R\sim 0.5$, signaling the break-down of perturbation
theory: the decay of a particle with mass $m_X\gg 10^6$~GeV, even if
coupled only to leptons, initiates a
cascade, very similar to that known in QCD.

For the quantitative study of EWLB cascades, we consider a broken SU(2)
gauge theory as a simplified model for the 
electroweak sector of the Standard Model (SM). Its particle content is
a triplet of gauge bosons $W^A_\mu$ with mass $m_W$,  
a physical Higgs $h$ with mass $m_h$ and $n_g=3$ generations of 
leptons $f=(l_L,\nu_{l,L})$, $l_R$, with $l=e$, $\mu$, $\tau$.
The SU(2) doublet $f$ is coupled to the gauge sector as 
$g_2 t^A_{bc}\bar f_b\gamma^\mu \frac{1}{2}(1-\gamma^5)f_c W_\mu^A$.

Since the terms generating the logarithms in Eq.~(\ref{R}) are
associated with collinear and/or soft emission of additional particles,
we introduce splitting functions $P_{ijk}(z)$, viz.
\be     
 \frac{d\Gamma_{n+1}}{d\Gamma_{n}}  \ap d\omega_{i\to jk} = 
 \frac{\alpha_2}{2\pi} \frac{dt}{t} dz P_{ijk}(z) \,,
\ee
where $d\Gamma_n$ is the decay width into $n$ particles with
virtuality $t$, $z$ is the energy fraction of the additional particle,
and $d\omega_{i\to jk}$ approximates $d\Gamma_{n+1}/d\Gamma_{n}$ in the
high-energy, small-angle scattering limit.

In the following, it is useful to
distinguish between transverse and longitudinal modes of the gauge bosons,
$W_T$ and $W_L$. Going from QCD to our broken SU(2) model,
there are only trivial changes for the transverse modes due to the
different structure constants and the V--A coupling,
\ba
&& P_{W_T W_TW_T} = 2 \left[ \frac{1-z}{z}+\frac{z}{1-z}+z(1-z)
\right] ,
\\
&& P_{W_T ff} = \frac{n_g}{4}\, [z^2+(1-z)^2] ,
\\
&& P_{f fW_T} =  
   \frac{3}{2}\,P_{f f' W_T^\pm}= 3 P_{f fW_T^0}=
\frac{3}{4} \,\frac{1+z^2}{1-z}.
\ea
%

External longitudinal modes $W_L$ can be replaced by the corresponding  
Goldstone bosons $\chi_W$ in $S$-matrix elements for $s\gg
m_W^2$~\cite{equiv}. Since $m_W$ acts as an infrared
cut-off and is assumed to be well above the experimental resolution, 
virtual corrections and real bremsstrahlung do not mix. Replacing
$W_L$ by the scalars $\chi_W$ makes clear that the splitting
functions involving $W_L$ do not contain terms singular in $z$. The
longitudinal modes $W_L$ are therefore subdominant compared to $W_T$
and we will neglect them.

Finally, we discuss whether the Higgs participates actively in the
cascade evolution. The Higgs can split into pairs of fermions or
gauge bosons. In the first case, all Yukawa couplings except the one to
the top quark $t$ can be neglected; moreover collinear singularities are
absent. Therefore, only the processes $t\to th$ and $h\to tt$  
have a minor importance at $Q^2$ close to the GUT scale.
In the second case, {\em i.e.} when splittings involve Higgses and gauge
bosons, we cannot define splitting functions in the conventional
sense. Instead, 
$d\omega_{h\to W_L W_L}$ and $d\omega_{h\to W_T W_T}$ 
are proportional to $g^2/t^2$.
The resulting branching probability is negligible.

Hence, the EW cascade is supported essentially by the same
splitting modes as in the unbroken theory. The proof of the factorization
of both collinear and soft singularities in the EW theory
follows the same logic 
as in QCD~\cite{DGLAP1,DGLAP2} and the effects of the
spontaneous symmetry breaking result at $s\gg m_W^2$ only in
subleading corrections~\cite{pozzor}. Thus one obtains the EW
evolution equations in the Altarelli-Parisi form~\cite{CCC} and is
able to establish the SU(2) charge coherence picture in the usual
way~\cite{khose}. 
The evolution equations in integral form including coherence effects
are
\ba
\label{D}
\lefteqn{
 D_{i\to j} (Q^2,x,t) =  \Delta_s^i(Q^2,t)\:\delta_i^j \:\delta(1-x) \:+}
\\ &&
+\; \sum_{kl}  \int_{t}^{Q^2}\!\frac{dt'}{t'} \int_x^{1-\epsilon _l(t')}
 \frac{dz}{2\pi z}\; \Delta_s^i(Q^2,t') \times  
\nonumber\\ && 
\times  \alpha_2[z^2(1-z)^2t'] \; P_{ikl}(z)
\;  D_{k\rightarrow j}(x/z,z^2t')
  \,,
\nonumber
\ea
where $D_{i\to j}(Q^2,x,t)$ defines the energy distribution 
of particles $j$ with energy fraction $x$ at scale $t$ produced 
by the parent particle
$i$ at scale $Q^2$. Here, $\epsilon _l(t')\ap m_l/\sqrt{t}$,
where $m_l$ is the mass of the particle $l$, $t=q^2/[z(1-z)]$, 
$q^2$  the virtuality of the parent particle and $z$ 
the energy fraction of the produced one. 
We use
$\alpha_2(t)=$
\mbox{$\alpha_2(m_Z^2)/[1+(b/4\pi)\ln(t/m_Z^2)\alpha_2(m_Z^2)]$} 
with $b=19/6$  for one Higgs doublet and $\alpha_2(m_Z^2)\sim 1/29.6$.
%

In Eq.~(\ref{D}), we have introduced the Sudakov form factors
$\Delta_s^i(Q^2,t)$ 
giving the probability of no branching for the parton of type $i$
in the scale range between $Q^2$ and $t$,
\ba
 \ln\Delta_s^i(Q^2,t)& = & -\sum_{kl}\int_{t}^{Q^2}\!\!\frac{dt'}{t'}
    \int_{\epsilon_k(t')}^{1-\epsilon_l(t')} \! 
\label{S}
\\ &\times & 
\frac{dz}{2\pi}  \: \alpha_2[z^2(1-z)^2t']  \, P_{ikl}(z)  \,.
\nonumber
\ea

Equations (\ref{D}-\ref{S}) are very similar to the analogous ones
in QCD and allow to use the probabilistic scheme of Ref.~\cite{Ma84}.
There is however an important difference between the EWLB and QCD
cascades  
in the physical meaning of the cut-off value $\epsilon_k(t')$. 
In QCD, it is defined by the scale $q^2_{\min}$
at which the perturbative evolution of the cascade is terminated, 
$\epsilon_{\rm QCD}(t')=\sqrt{q^2_{\min}/t'}$.
By contrast, $q^2_{\min}$ of the EWLB cascade is given by the
physical masses of the particles produced in the splitting $i\to kl$.
For $W_T\to ff$, $\epsilon_k(t')\ap 0$, i.e., it is
much smaller than the one in $W_T\to W_TW_T$, where 
$\epsilon_k(t')\ap m_W/\sqrt{t'}$. As a
consequence, the $W_T$ does not dominate the cascade evolution as 
strongly as gluons in QCD and the number of $W_T$ leaving the cascade is 
smaller than the number of leptons.

The EWLB cascade does not terminate abruptly at a certain $Q^2$.
Instead, the probability for no further
branching increases smoothly from $\Delta_s^l(Q^2,0)=0$ for 
$Q^2\to\infty$ to $\Delta_s^l(0,0)=1$. Nevertheless, one can define
$Q^2_{\rm cr}$  by $\Delta_s^l(Q^2_c,0) = 0.5$, i.e. by that
virtuality $Q^2_{\rm cr}$ at which the probabilities for further
branching and for non-branching become equal. Then $Q_{\rm cr}$ is
given by $Q_{\rm cr}\sim 10^6$~GeV and the probability for further
splittings decreases indeed drastically for lower $Q$.

\begin{figure}[h]
\begin{picture}(7,6.4)
\put(-0.2,.7) {
  \epsfig{file=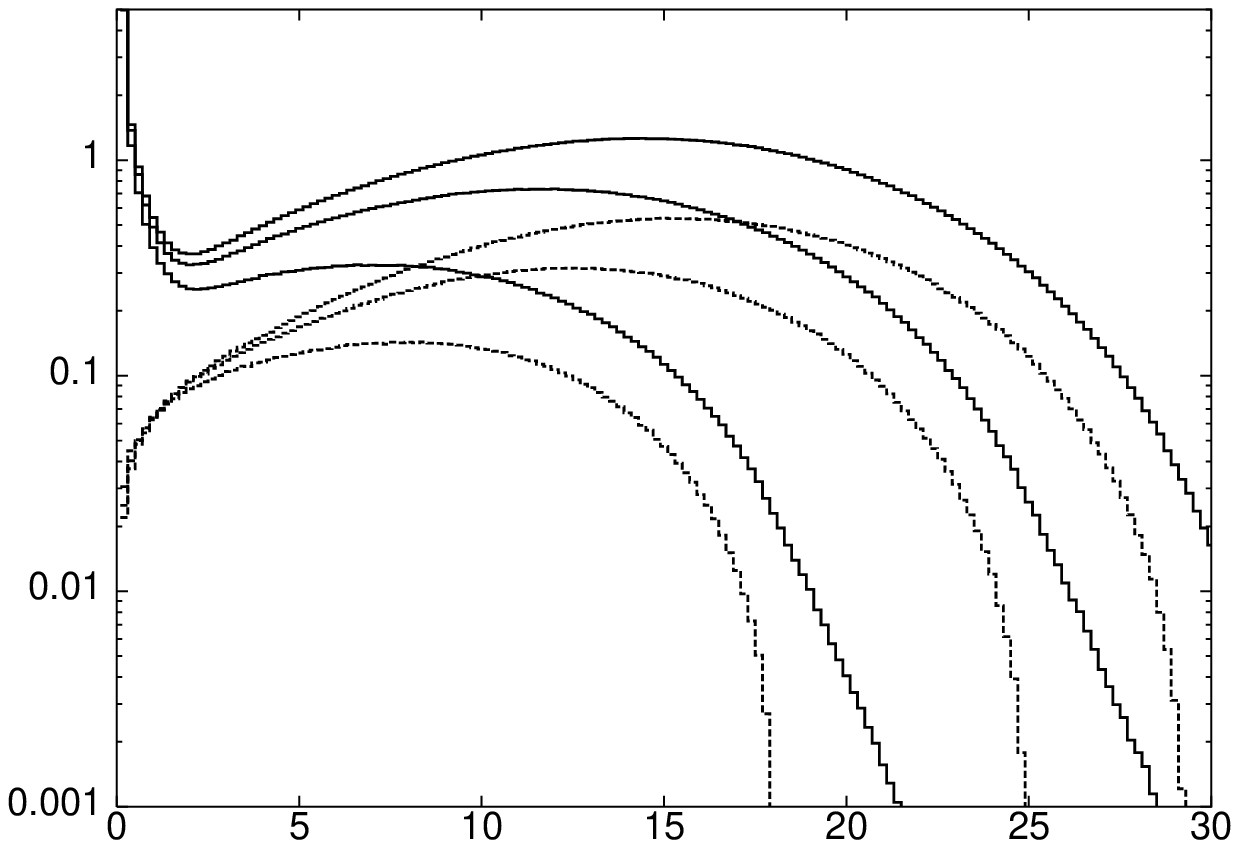,height=5.8cm,width=7.4cm,angle=0} }
\put(3.0,0.3) {$l=\ln(1/x)$}
\put(-0.6,3.7) {{\Large $\frac{dN}{dl}$}}
\put(1.1,4.9) {$f$}
\put(1.,3.2) {$W$}
\end{picture}
\caption{\label{f} 
Energy spectra $dN/dl$ of leptons $f$ (solid lines)
and $W$'s (broken lines) produced
in $X\to \bar\nu\nu\to$~all; for $m_X=10^{16}$,
$10^{13}$ and $10^{10}$~GeV from top to down.}
\end{figure}

We choose to solve the evolution equation with a Monte Carlo simulation. 
This method gives us the advantage of including non-abelian
charge coherence effects through angular ordering.
In Fig.~\ref{f}, we show the spectra $dN/dl$, where $l=\ln(1/x)$ and
$x=2E/m_X$, of leptons and $W$'s for $m_X=10^{10}$,
$10^{13}$ and $10^{16}$~GeV and for the decay mode $X\to \bar\nu\nu\to$~all.
The average number of produced leptons and $W$'s increases from 9 for 
$m_X=10^{10}$~GeV to 47 for $m_X=10^{16}$~GeV, where we consider the
$\mu^\pm$, $\tau^\pm$ and the $W^a$ as stable particles. 
The first bin contains between $7\%$ ($m_X=10^{10}$~GeV) and
$2\promille$ ($m_X=10^{16}$~GeV) of prompt neutrinos with $E=m_X/2$ from
no-cascading decays. 
Leptons which stop branching after the first splitting produce a tail 
superimposed to the usual Gaussian, clearly visible for $l\lsim 3$.

\unitlength1.0cm

Let us discuss now the inclusion of QCD partons in our model.
Since the EW gauge bosons split also into quarks $q$,
there will be a mutual transmutation of ``leptonic'' and
``QCD'' cascades. The shape of the hadron energy spectra is however
only marginally influenced by the leptons, first because the QCD
cascade is determined mainly by gluons $g$ and second because the
probability of $q\to q+g$ is much larger than of $q\to q+W$. 
On the other hand, the splittings $W_T\to qq$ act
continuously as a sink for the particles and energy of the EWLB
cascade. These splittings are taken into account in the MC
simulation and the spectra shown in Fig.~1.

We have calculated also (see Table 1) a quantity interesting for
decays of $X$-particles in the Universe: the fraction of energy 
$f_{\rm em}$, transferred to electrons and photons in the
decay mode $X\to\bar\nu\nu$ as function of $m_X$. These
particles initiate electromagnetic (e-m) 
cascades in the intergalactic space, through 
interactions with microwave (CMB) photons. 
In the calculation of $f_{\rm em}$, we assumed that all hadrons
produced in $W_T\to q\bar{q}\to$ hadrons are pions, while we used the SM
branching ratios for $\mu$ and $\tau$ decays.

%
The above calculations have interesting consequences for Ultra High 
Energy Cosmic Rays (UHECR) produced by superheavy Dark
Matter (DM)~\cite{SHDMCR} and, in particular, by $Z$-bursts~\cite{Z}.
In the latter model, UHECR 
are produced through the resonant production of $Z$-bosons in the
collisions of UHE neutrinos with DM neutrinos, $\nu+\bar\nu \to
Z^0$. The resonant energy of UHE neutrino is 
$E_0=m_Z^2/2m_{\nu}=4.2\times10^{21}m_{\rm eV}^{-1}$~eV, where $m_{\rm eV}$ 
is the mass of the DM neutrino in eV.

Decays of these $Z$-bosons produce UHECR which energy spectrum,
according to recent calculations
\cite{FKR,KKSS}, has a weak Greisen-Zatsepin-Kuzmin 
cutoff and explains well the observations. 
A remarkable feature of this model is that the production of UHE
particles with energies higher than $10^{20}$~eV does not involve exotic 
elementary particle physics, while its drawback consists in the
necessity of an enormous flux of resonant neutrinos. 

A widely discussed possibility is the production 
of UHE neutrinos in astrophysical sources. This is an unrealistic
option because it requires too high luminosities of 
the astrophysical sources: we can estimate the required flux of resonant
neutrinos $I_{\nu}(E_0)$ and therefrom the neutrino energy density
$\omega_{\nu}$ as  
$\omega_{\nu} \approx (2.4 - 3.6)\times 10^{-13} m_{\rm eV}^{-0.5}$~erg/cm$^3$.
In these calculations, we have used that for the flat spectra 
generated by  heavy particle decays, such as $Z$-bosons and $X$-particles, 
UHE photons dominate at the highest energies $E \geq 1\times 10^{20}$~eV
\cite{BBV}.
The resulting neutrino  luminosity of a source, estimated as 
$L_{\nu} \sim \omega_{\nu}/(n_st_0)$, where $n_s$ is the source density 
and $t_0$ the  age of the Universe, is too high:
$(8 -12)\times 10^{44}$~erg/s, if the sources are normal galaxies, and 
$(8 -12)\times 10^{46}$~erg/s in the case of Seyfert galaxies. Our
result confirms similar conclusions of Ref.~\cite{KKSS}, reached
on the basis of a limit due to the diffuse $\gamma$-ray background.  

Alternatively,
$Z$-burst neutrinos can be generated by decays of superheavy particles, 
either existing as DM particles or generated by topological defects. 
In the usual models this possibility is ruled out by the diffuse
$\gamma$-ray 
background~\cite{BV}. However, in Ref.~\cite{GK2} it is suggested that
this limit can be evaded, if X-particles decay exclusively to
neutrinos, i.e. $X \to \bar\nu\nu$. 

We shall demonstrate now that electroweak cascading rules out 
this last hope, too. 

EW cascading modifies the e-m cascading upper limit for UHE
neutrinos~\cite{BV,baksan}  
by the factor $f_{\nu}/f_{\rm em}$, where $f_{\nu}$ is the fraction of
energy $m_X$ transferred to neutrinos. 
Then the upper limit
on the UHE neutrino flux is given as 
\be
I_{\nu}(E)<\frac{c}{4\pi}\frac{f_{\nu}}{f_{\rm em}}
\frac{\omega_{\rm cas}}{E^2},
\label{bb-cas}
\ee
with $\omega_{\rm cas}\approx 2\times 10^{-6}$~eV/cm$^3$ according to
EGRET observations.
Taking $I_{\nu}(E_0)$ from Eq.~(\ref{bb-cas}), one can calculate 
the flux of UHE photons produced in $Z$-bursts. At $E\geq 1\times
10^{20}$~eV, this flux should be of the order of the observed UHECR flux.
With $f_{\rm em}=0.2$ from Table 1, we obtain 
$E^3I_{\gamma}(E) \approx (5 - 8)\times 10^{21}m_{\rm eV}$~eV$^2$/m$^2$s sr,
which is almost three orders of magnitude less than observed.

{\em Summary---}
EW cascading in decays of superheavy particles results in 
EWLB cascades similar to the cascade of QCD partons, 
if $m_X >10^{6}$~GeV.
Thereby,  the flux of prompt neutrinos, e.g., in annihilations of DM
particles is 
reduced.  
EW cascades allow a probabilistic interpretation
and exhibit destructive coherence at small $x$. 
The generation of electrons and photons, which are able to start e-m
cascades on the CMB, in the EW cascade allows to exclude 
those $Z$-burst models, in which the
resonant neutrinos are produced in $X\to\bar\nu\nu$ decays.
Thus, the production of resonant neutrinos in astrophysical sources 
and by decays of superheavy particles (both 
as DM particles or produced by topological defects) result in 
neutrino fluxes too low for the $Z$-burst model. The only possibility 
left is the oscillation of sterile neutrinos, e.g. in
hidden-sector/mirror models~\cite{BV},  into ordinary ones.

The work of VB was supported in part by INTAS (grant 99-01065).
MK acknowledges an Emmy Noether--fellowship of the Deutsche
Forschungs\-ge\-meinschaft (DFG) and SO support by the German Ministry
for Education and Research (BMBF).

\begin{table}
\caption{\label{tab1} 
Energy fraction $f_{\rm em}$ transferred to electron and photons  
in the decay $X\to\bar\nu\nu$ of a particle with mass $m_X$.}
\begin{tabular}{c||c|c|c|c|c|c}
$m_X$/GeV       & $10^6$ & $10^8$ & $10^{10}$ & $10^{12}$&  $10^{14}$ & $10^{16}$  \\\hline
$f_{\rm em}$/\% & 11     &  15    &  17       & 19       & 20         & 20
\end{tabular}
\vskip0.3cm
\end{table}

\widetext


\begin{thebibliography}{00}

\vspace*{-0.7cm}
\bibitem{EW}
P.~Ciafaloni and D.~Comelli,
Phys.\ Lett.\ B {\bf 446}, 278 (1999);
M.~Beccaria {\it et al.},
Phys.\ Rev.\ D {\bf 61}, 073005 (2000);
V.~S.~Fadin {\it et al.}, 
Phys.\ Rev.\ D {\bf 61}, 094002 (2000);
W.~Beenakker and A.~Werthenbach,
Phys.\ Lett.\ B {\bf 489}, 148 (2000);
M.~Hori, H.~Kawamura and J.~Kodaira,
Phys.\ Lett.\ B {\bf 491}, 275 (2000);
A.~Denner and S.~Pozzorini,
Eur.\ Phys.\ J.\ C {\bf 18}, 461 (2001);
J.~H.~K\"uhn {\it et al.}, 
Nucl.\ Phys.\ B {\bf 616}, 286 (2001).
For early works see
W.~Beenakker {\it et al.},
Nucl. Phys. {\bf B410}, 245 (1993),
Phys. Lett. {\bf B317}, 622 (1993).
For a recent review see
M.~Melles, hep-ph/0104232.

\bibitem{DGLAP1} 
V.~N.~Gribov and L.~N.~Lipatov, Sov. J. Nucl. Phys. {\bf 15}, 438 
and 675 (1972).  

\bibitem{DGLAP2} 
G.~Altarelli and G.~Parisi, Nucl. Phys. {\bf B126}, 298 (1977);
Yu.~L.~Dokshitzer, Sov. Phys. JETP {\bf 46}, 641 (1977).  

\bibitem{CCC} 
M.~Ciafaloni, P.~Ciafaloni, D.~Comelli,
Phys.\ Rev.\ Lett.\  {\bf 88}, 102001 (2002).

\bibitem{wegr}
S.~Weinberg,
Phys.\ Rev.\  {\bf 140}, B516 (1965);
J.~R.~Ellis, M.~K.~Gaillard and G.~G.~Ross,
Nucl.\ Phys.\ B {\bf 111}, 253 (1976).



\bibitem{equiv}
J.~M.~Cornwall, D.~N.~Levine and G.~Tiktopoulos, Phys. Rev. {\bf D10}, 1145
(1974); 
B.~W.~Lee, C.~Quigg and H.~B.~Thacker,
Phys.\ Rev.\ D {\bf 16}, 1519 (1977);
M.~S.~Chanowitz and M.~K.~Gaillard,
Nucl.\ Phys.\ B {\bf 261}, 379 (1985).

\bibitem{pozzor} 
S.~Pozzorini, hep-ph/0201077 and references therein.

\bibitem{khose} 
A.~Bassetto, M.~Ciafaloni and G.~Marchesini,
Phys.\ Rept.\  {\bf 100}, 201 (1983);
Yu.~L. Dokshitzer {\it et al.}, {\em Basics of Perturbative QCD\/},
Editions Fronti{\`e}res 1991. 


\bibitem{Ma84}
G.~Marchesini and B.~R.~Webber,
Nucl. Phys. {\bf B238}, 1 (1984).

\bibitem{SHDMCR}
V.~Berezinsky, M.~Kachelrie\ss, and A.~Vilenkin, Phys. Rev. Lett. {\bf
79}, 4302 (1997);
V.~A.~Kuzmin and V.~A.~Rubakov, Phys. Atom. Nucl. {\bf 61}, 1028 (1998).

\bibitem{Z}
D.~Fargion, B.~Mele and A.~Salis,
Astrophys. J. {\bf 517}, 725 (1999);
T.J.~Weiler,
Astropart. Phys.  {\bf 11}, 303 (1999).



\bibitem{FKR}
Z.~Fodor, S.~D.~Katz and A.~Ringwald,
hep-ph/0105064
and
hep-ph/0203198.
%
See also S.~Yoshida, G.~Sigl and S.~J.~Lee,
Phys.\ Rev.\ Lett.\  {\bf 81}, 5505 (1998).


\bibitem{KKSS}
O.~E.~Kalashev, V.~A.~Kuzmin, D.~V.~Semikoz and G.~Sigl,
hep-ph/0112351.


\bibitem{GK2}
G.~Gelmini and A.~Kusenko,
Phys. Rev. Lett. {\bf 84}, 1378 (2000).

\bibitem{BBV}
V.~Berezinsky, P.~Blasi and A.~Vilenkin,
Phys.\ Rev.\ D {\bf 58}, 103515 (1998).

\bibitem{BV}
V.~S.~Berezinsky and A.~Vilenkin,
Phys.\ Rev.\ D {\bf 62}, 083512 (2000).


\bibitem{baksan}
V.~S.~Berezinsky, in Proc. of ``Neutrino-77'', Baksan , USSR, ed. M.A.Markov 
{\bf 1}, 177 (1977).



\end{thebibliography}
\end{document}